# Two-dimensional Ca-Cl crystals under ambient conditions observed directly by cryo-electron microscopy

Lei Zhang[1]*, Guosheng Shi[2]*, Bingquan Peng[1,3]*, Pengfei Gao[1]*, Ni Zhong[4], Liuhua Mu[3], Han Han[3], Liang Chen[5], Lijuan Zhang[3], Peng Zhang[1], Lu Gou[6], Yimin Zhao[1], Shanshan Liang[3], Chungang Duan[4], Long Yan[3]#, Shengli Zhang[1]# & Haiping Fang[3]#

[1]MOE Key Laboratory for Nonequilibrium Synthesis and Modulation of Condensed Matter, School of Science, Xi'an Jiaotong University, Xi'an 710049, China
[2]Shanghai Applied Radiation Institute, Shanghai University, Shanghai 200444, China
[3]Shanghai Institute of Applied Physics, Chinese Academy of Sciences, Shanghai 201800, China
[4]Key Laboratory of Polar Materials and Devices, Ministry of Education, East China Normal University, Shanghai 200241, China
[5]Zhejiang Provincial Key Laboratory of Chemical Utilization of Forestry Biomass, Zhejiang A&F University, Lin'an, Zhejiang 311300, China
[6]Department of Applied Physics and State Key Laboratory for Manufacturing Systems Engineering, Xi'an Jiaotong University, Xi'an 710049, China

*These authors contributed equally to this work.
#Corresponding author. E-mail: fanghaiping@sinap.ac.cn (H.F.); zhangsl@xjtu.edu.cn (S.Z.); yanlong@sinap.ac.cn (L.Y.)

**Recently, we report the direct observation, under ambient conditions, of $Na_2Cl$ and $Na_3Cl$ as two-dimensional (2D) Na–Cl crystals, together with regular NaCl, on reduced graphene oxide membranes and on the surfaces of natural graphite powders from salt solutions far below the saturated concentration[1]. However, what are these abnormal stoichiometries for high valence ions, such as calcium ions[2-5] and copper ions still remain unknown. Here, using cryo-electron microscopy, we report the direct observation of two-dimensional (2D) Ca-Cl crystals on reduced graphene oxide (rGO) membranes, in which the calcium ions are only monovalent (i.e. +1). Remarkably, metallic properties rather than insulating are displayed by those CaCl crystals. We note that such CaCl crystals are obtained by simply incubating rGO membranes in salt solutions below the saturated concentration, under ambient conditions. Theoretical studies show that the formation of those abnormal crystals is attributed to the strong cation–π interactions of the $Ca^{2+}$ ions with the aromatic rings in the graphitic surfaces. Since strong cation–π interactions also exist between other metal ions (such as $Mg^{2+}$, $Fe^{2+}$, $Co^{2+}$, $Cu^{2+}$, $Cd^{2+}$, $Cr^{2+}$ and $Pb^{2+}$) and graphitic surfaces[6], similar 2D crystals with abnormal valence state of the metal cations and corresponding abnormal properties are highly expected. The 2D crystals with monovalent calcium ions show unusual electronic properties, and can be applied in catalyzer, hydrogen storage, high-performance conducting electrodes and sensors. These findings also produce functionalized graphene including compact "graphene—metallic CaCl—insulating $CaCl_2$" junction that can serve as transistors down to the atomic scale, and other devices for magnetic, optical and mechanical applications.**

C is one of the most important element for human beings. Since the discovery of $C_{60}$[7], carbon nanotubes[8] and graphene[9], carbon continuously produces surprises. However, until recently, most of the studies focus on the morphologies and properties of those carbon-based structures and their complexes with other components having normal stoichiometry. Here, we report metallic 2D Ca-Cl crystals with monovalent (i.e. +1) calcium ions on rGO membranes under ambient conditions observed directly by cryo-electron microscopy.

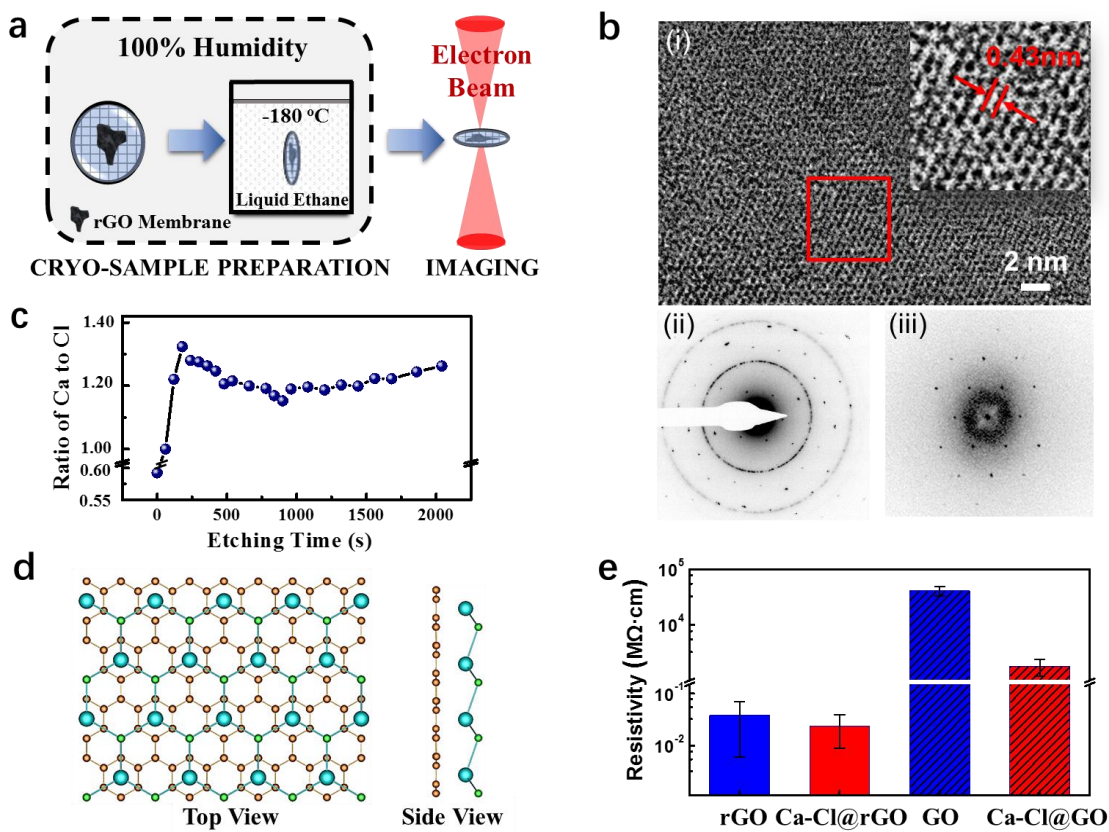

**Figure 1 | Two-dimensional (2D) Ca-Cl crystals from salt solution below the saturated concentration. a**, Schematic drawings of the sample preparation processes. **b**, (i) Cryo-EM image of the Ca-Cl crystals at the edge of the reduced graphene oxide (rGO) membrane. Inset shows a magnified image of the area outlined in red with a lattice spacing ~ 0.43 nm. (ii) Diffraction pattern of a typical crystal structure by cryo-EM in electron diffraction mode, showing a hexagonal lattice pattern with six first-order maxima points at ~1/0.43 nm$^{-1}$. (iii) Fourier transform of the entire bright-field image showing the same hexagonal lattice as in (ii). **c**, Atomic ratio of Ca to Cl as a function of the etching time measured by XPS during etching by argon ion for a sample of *dried* rGO membrane with Ca and Cl. The *dried* membrane was obtained by first incubating in 5.0 M CaCl$_2$ solutions overnight and then centrifuging to remove the free solution and drying at 70 °C for 4 hours. **d**, One stable structure of CaCl crystal modules adsorbed on graphene sheet from theoretical computations. Ca, Cl and C atoms are in blue, green and brown, respectively. **e**, Electrical resistivity measured by using the multimeter with two electrodes connecting with the up and down surfaces of the *dried* rGO and GO membranes, respectively.

The 2D Ca-Cl crystals are obtained by soaking rGO membranes in CaCl$_2$ solution below the saturated concentration under ambient conditions. Freestanding rGO membranes[1,10,11] were prepared from an rGO suspension via the drop-casting method as described in ref. 8 (see PS1 of Supplementary Information, SI). These membranes were then incubated in 5.0 mol/L (M) CaCl$_2$ solution overnight under ambient conditions. This concentration is ~ 75% of the saturation concentration for CaCl$_2$ solution at 20 °C. The membranes with salt solution were then split into small pieces and analyzed by cryo-electron microscopy (cryo-EM)[12,13]. During sample preparation process, environmental parameters were set at 20°C with 100% humidity, which could keep a thin film of solution with original salt concentration coating rGO membranes due to flash-freezing procedure, as demonstrated by many other cryo-EM studies on biological macromolecules and inorganic complexes[14-16] (Fig. 1a). As a result, at the edge of the rGO membrane relatively thin (few layers of rGO sheets) membranes coated by thin amorphous ice could be found (Fig. S1).

Within several such regions, stable single-crystal diffraction patterns could be observed directly even after rather high electron dose exposure, which made amorphous ice at the top melt and moving (Movie S1). Typical high-resolution cryo-EM images of such crystals in rGO membranes are shown in Fig. 1b. These high-contrast images (Fig. 1bi) show that such crystals have a lattice spacing of 4.40 ± 0.17 Å, corresponding to a graphene-like honeycomb lattice with a side length of 2.93 ± 0.12 Å. Further, electron diffraction and fast Fourier transform analyses[17] of the Ca-Cl lattice yielded a hexagonal lattice with first-order maximal points at $(1 \pm 0.04)/4.40$ Å$^{-1}$ (Figs. 1bii,1biii, and S2). The high-resolution TEM images and their Fourier transform patterns of the same crystals after tilting at 20° - 40° (Fig. S3) show that such crystals are relatively thin (< 1nm) and are 2D crystals.

In order to explore the elemental composition of these crystals, we used X-ray photoelectron spectroscopy (XPS) (see PS1 of SI) to measure the atomic ratio of Ca to Cl in *dried* rGO membranes with Ca and Cl. The *dried* membrane with Ca and Cl was obtained by first incubating in 5.0 M CaCl$_2$ solutions overnight and then centrifuging to remove free solution and drying at 70 °C for 4 hours. By etching with argon ions, the depth profile shows that the Ca:Cl ratios vary from a value below 0.6:1 at the membrane top surface to a stable ~1.2:1 in inner sheets (Fig. 1c). The membrane surface may have regular CaCl$_2$ due to evaporation of adsorbing salt solution, but at inner sheets there are mainly Ca-Cl crystals with a stable Ca:Cl ratio of ~1:1. The existence of elemental ratio of Ca to Cl close to 0.5:1 was further confirmed by conventional (non-cryogenic) TEM EDS analysis (see PS1 and PS4 of SI). We note that the same CaCl crystals as shown in Fig. 1b were also observed by conventional TEM in those dried samples (Fig. S5).

Now we discuss the possible structures of the crystals. Using density-functional theory (DFT)[18,19] (see PS1 and PS5 of SI), we obtained a stable structure with a Ca:Cl ratio of 1:1, which could be denoted as CaCl (Figs. 1d and S10a, Model I). In this structure, the Ca occupies a plane parallel to the graphene sheet and is located in the center of nonadjacent aromatic rings. Each Ca is surrounded by three Cl with a horizontal distance equals to twice the C-C bond length in graphene, consistent with the crystal structure revealed by cryo-EM observations (Fig. S11a). Difference charge density analyses (Fig. S12) revealed significant accumulation of the valence electrons of Ca and the π charges of aromatic ring on graphene sheet, showing strong cation-π interactions (see PS5 of SI). Further numerical Bader charge analysis[20] shows that Ca loses ~1.5 e$^-$, in which about 0.85 e$^-$ transfers to Cl and the rest to graphene (Table S2). Such charge transfer helps to stabilize the structure of CaCl on graphene sheet. DFT computations have also shown that there is stable 2D crystal structure between two graphene sheets with an appropriate distance (Model II), while this structure has the same Ca:Cl ratio and projection of crystal structure in graphene plane as Model I (Fig. S10). The only differences between Models I and II are the vertical locations (relative to the graphene sheet plane) of the Ca and Cl. Further, there are other possible structures of Ca-Cl (Fig. S10a), as well as regular $CaCl_2$, $Ca_3Cl_4$, $Ca_3Cl_2$ (Fig. S10b), on graphene surface or confined between graphene sheets.

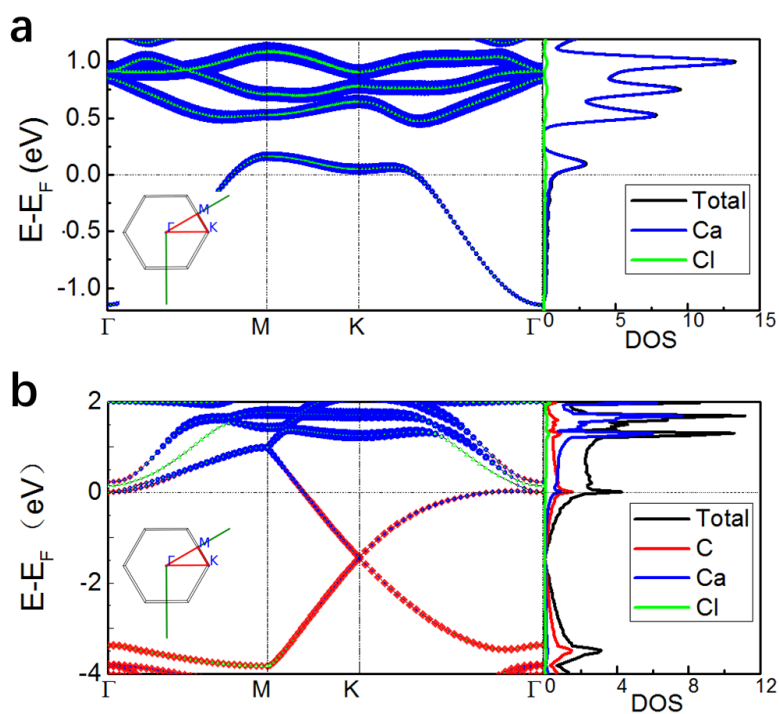

**Figure 2 | Electronic structure of Model I. a**, Projected band structure and projected DOS on each element in the primitive cell of CaCl alone. **b**, Projected band structure and projected DOS on each element in the primitive cell of full model. The size of the points in (a) and (b) indicates the weight of each element's contribution. Inset: the high symmetry q point path in the Brillouin Zone. $E_F$, Fermi energy.

These CaCl crystals with unusual electronic structure of monovalent (i.e. +1) calcium ions emerge revolutionary new properties for the crystals with calcium ions, such as metallicity. The projected electronic band structure and density of states (DOS) of Model I shows that CaCl crystals have distinct metallicity in which Ca plays a dominant role (Fig. 2a). The complexes of graphene with CaCl crystals also has strong metallicity, and that Ca and Cl together contribute ~37.7% of the total electronic states near the Fermi level, indicating that CaCl crystals are also metallic in this system (Fig. 2b). More remarkably, compared with pure 2D graphene and the 2D CaCl, this system shows greatly enhanced electrical conductivity along the horizontal and vertical directions (Figs. S13, S14 and Table S4).

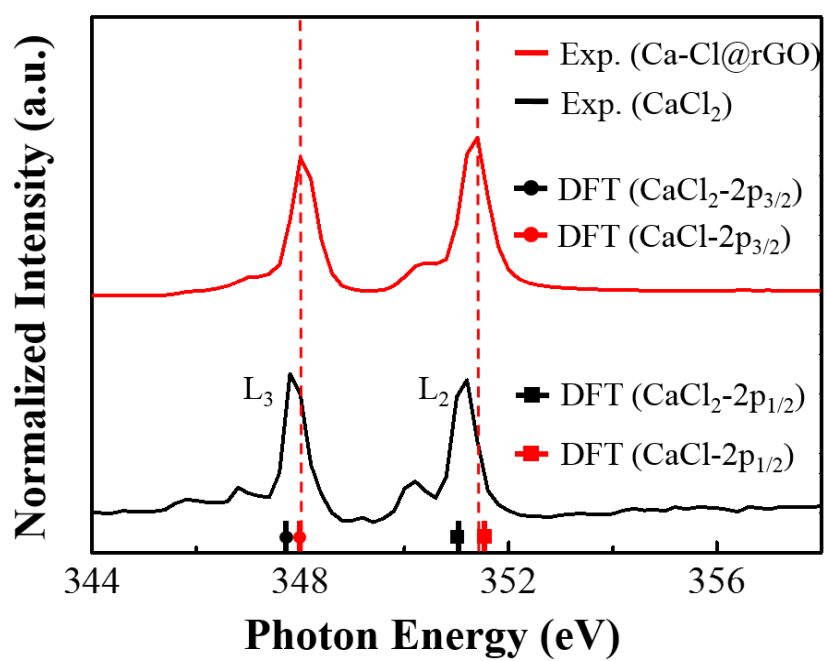

**Figure 3** | Calcium near-edge X-ray absorption fine structure spectra (NEXAFS) as well as the computation of the density function theory for calcium 2p core excitation energies of the $CaCl_2$ crystal and the Ca-Cl crystals in rGO membranes.

To confirm the abnormal valence state change, the Ca-Cl crystals in *dried* rGO membranes were analyzed by soft X-ray absorption spectroscopy (near-edge X-ray absorption fine structure spectra, NEXAFS). Fig. 3 and Fig. S19 show that, in addition to the peak of Ca $L_{2,3}$-edge for regular $CaCl_2$ at 347.80 eV, there is a new Ca $L_{2,3}$-edge peak at 348.00 eV. DFT computations of the Ca 2p core excitation energies of Ca-Cl crystals show that the new Ca $L_{2,3}$-edge peak is from the CaCl crystal of Model I (Fig. 3, see PS1 and PS6 of SI). This indicates that that there is valence electron change of the Ca ion from $Ca^{2+}$ to $Ca^{+}$ in these crystals.

Metallic properties of Ca-Cl crystals in rGO membranes have also been confirmed experimentally by measuring the electrical resistivity (see PS1 and PS8 of SI). Fig. 1e shows that the resistivities of pure rGO membranes and *dried* rGO membranes with Ca-Cl crystals are both less than ~ $4.0 \times 10^{-2}$ MΩ·cm, indicating their good conductivity. We note that our result on rGO membranes is consistent with metallic properties reported in the literature[21]. In contrast, for *dried* GO membranes with high resistivity, the average resistivity of *dried* GO membranes containing Ca-Cl crystals was reduced at least one order of magnitude (Fig. 1e). The electrical conductivities of the metallicity of Ca-Cl crystals have been further demonstrated by using conductive atomic force microscopy (AFM) on the graphene sheets (see PS1 and PS8 of SI).

The +1 valence in these CaCl crystals provides distinct adsorption capacity with other materials, which not only can fill the insufficient adsorption of metallic calcium atoms but also overcome the exorbitant absorption of +2 calcium ions. This provides many potential applications such as high-capacity hydrogen medium. DFT computations show that the maximal number of stably adsorbed $H_2$ on each Ca in a unit cell is 3 with adsorption energy of ~ 0.15 eV per $H_2$ up to 200 K (Figs. S25 and S26), while the hydrogen molecules are released above this temperature.

Nature continuously produces surprises, and our results demonstrate another one. 2D Na-Cl crystals with abnormal stoichiometries on graphite or graphene[1] and at extreme conditions, such as high pressure[22] have been reported. Here, we not only using cryo-electron microscopy observed 2D CaCl crystals with unconventional stoichiometries directly on reduced rGO membranes in unsaturated solution but also experimentally demonstrate the +1 valence of Ca cations and the metallic property of the crystals. The monovalent calcium ions show unusual electronic properties, which can be applied in catalyzer, hydrogen storage, high-performance conducting electrodes and sensors. Further, the metallic property of the crystals on graphene produces functionalized graphene that can serve as transistors with enhanced electronic conducting capacity down to the atomic scale, and devices for other potential magnetic, optical and mechanical applications. For example, compact "graphene—metallic CaCl" could be directly obtained by incubating rGO membranes with unsaturated $CaCl_2$ solution. "Metallic CaCl—insulating $CaCl_2$" and "graphene—metallic CaCl—insulating $CaCl_2$" junctions could be obtained by growing layers of regular $CaCl_2$ crystals on the 2D metallic CaCl crystals after further immersing the *dried* rGO membranes in saturated $CaCl_2$ solution.

Such abnormal Ca-Cl crystals are attributed to the strong cation–π interactions between $Ca^{2+}$ ions and aromatic rings in the graphene surfaces. We expect similar abnormal crystals for other metal cations, since there are also strong cation–π interactions exist between other metal ions (such as $Mg^{2+}$, $Fe^{2+}$, $Co^{2+}$, $Cu^{2+}$, $Cd^{2+}$, $Cr^{2+}$ and $Pb^{2+}$) and graphitic surfaces[1,6,23-29]. In fact, our NEXAFS and XPS experiments show, under ambient conditions, Cu-Cl crystals with +1 copper ions of long life time, which can stably exist for several days (see PS7 of SI). This would help to overcome the difficulty of the short life time in the application as catalyzers for the +1 copper ion[30]. Further, considering the wide distribution of metallic cations and carbon on earth, such nanoscale "special" compounds with previously unrecognized properties may be ubiquitous in nature.

## ACKNOWLEDGMENTS

We thank Dr. Philip Ball for his constructive suggestions. The supports from the National Natural Science Foundation of China (Nos. 11774279, 11574339, and 11774280), the National Science Fund for Outstanding Young Scholars (No. 11722548), the Key Research Program of Frontier Sciences of the Chinese Academy of Sciences (No. QYZDJ-SSW-SLH053), the Key Research Program of the Chinese Academy of Sciences (No. KJZD-EW-M03), Shanghai Supercomputer Center of China and the Special Program for Applied Research on SuperComputation of the NSFC-Guangdong Joint Fund (the second phase), and Young Talent Support Plan of Xi'an Jiaotong University are acknowledged. We also thank the Instrument Analysis Center of Xi'an Jiaotong University for cryo-EM imaging, and the beam line 08U1A of the Shanghai Synchrotron Radiation Facilities (SSRF) for the measurements of XANES near the Ca and Cu L edge.

**Supplementary Information:**
Materials and Methods
Supplementary Text
Figs. S1 to S26
Tables S1 to S4
References
Caption for Movie S1